\documentclass[pre, twocolumn, superscriptaddress, floatfix, nofootinbib]{revtex4-1}  

\usepackage[paperwidth=210mm,paperheight=297mm,centering,hmargin=2cm,vmargin=2.5cm]{geometry}

\usepackage[pdftex]{graphicx}
\usepackage{amsmath,amsfonts,amssymb}
\usepackage{natbib}
\usepackage{MnSymbol}
\usepackage{csquotes}
\usepackage{lipsum}

\usepackage{hyperref}
\usepackage{xcolor} 
\usepackage{array}
\usepackage{pgfplots}
\usepackage{graphicx}
\usepackage{dsfont}

\definecolor{bluemoi}{rgb}{0.25,0.50 ,0.75} 

\hypersetup{
	bookmarksopen=true,
	pdftitle=" ",
	pdfauthor=" ", 
	pdfsubject=" ", 
	pdftoolbar=true, 
	pdfmenubar=true, 
	pdfhighlight=/O, 
	colorlinks=true, 
	pdfpagemode=UseNone, 
	pdfpagelayout=SinglePage, 
	pdffitwindow=true, 
	linkcolor=bluemoi, 
	citecolor=bluemoi, 
	urlcolor=bluemoi 
}

\makeatletter
\renewcommand{\figurename}{\sf \textbf{Figure}}
\renewcommand{\thefigure}{\arabic{figure}}
\renewcommand{\fnum@figure}{\sf\textbf{\figurename}~\textbf{\thefigure}}
\renewcommand{\tablename}{\sf\textbf{Table}}
\renewcommand{\thetable}{\arabic{table}}
\renewcommand{\fnum@table}{\sf\textbf{\tablename}~\textbf{\thetable}}
\makeatother

\begin{document}
	
\title{Uncovering the socioeconomic structure of spatial and social interactions in cities} 
	
\author{Maxime Lenormand}
\thanks{Corresponding authors: maxime.lenormand@inrae.fr \& horacio@ecoinformatica.cl who contributed equally to this work.}
\affiliation{TETIS, Univ Montpellier, AgroParisTech, Cirad, CNRS, INRAE, Montpellier, France}
	
\author{Horacio Samaniego}
\thanks{Corresponding authors: maxime.lenormand@inrae.fr \& horacio@ecoinformatica.cl who contributed equally to this work.}
\affiliation{Laboratorio de Ecoinform\'atica, Instituto de Conservación, Biodiversidad y Territorio, Universidad Austral de Chile, Campus Isla Teja s/n, Valdivia, Chile}

\begin{abstract} 
The relationship between urban mobility, social networks and socioeconomic status is complex and difficult to apprehend, notably due to the lack of data. Here we use mobile phone data to analyze the socioeconomic structure of spatial and social interaction in the Chilean's urban system. Based on the concept of spatial and social events, we develop a methodology to assess the level of spatial and social interactions between locations according to their socioeconomic status. We demonstrate that people with the same socioeconomic status preferentially interact with locations and people with a similar socioeconomic status. We also show that this proximity varies similarly for both spatial and social interactions during the course of the week. Finally, we highlight that these preferential interactions appear to be holding when considering city-city interactions.
\end{abstract}
	
\maketitle
	
\section*{Introduction}
	
Securing equal opportunities to access public infrastructure is a major challenge in urban planning \cite{Hall2019}. More so, given the large concentration of wealth observed among the increasingly urban economies worldwide \cite{Alvaredo2018}. While these issues have largely been discussed in transportation \cite{El2016}, sociology \cite{Jones2014}, and physics \cite{Lenormand2020a} among other disciplines, the current deluge of spatially contextual information regarding the mobility and social interaction among humans is offering precise quantitative descriptions of emerging patterns of spatial socio-economic mixing across cities \cite{Steele2017,Dannemann2018,Cottineau2019,Alessandretti2020,Lenormand2020a}.

The analysis of trace information generated by mobile phones, credit cards and transit cards, among others, has shown to provide a simple, synoptic and near real-time descriptions of urban mobility that has expanded our understanding mobility strategies towards fine-grained and contextual representations of how travel budgets are segmented across the different dimensions of human life \cite{DeMontjoye2013,Blondel2015,Barbosa2018,Lenormand2016a,Lenormand2020b}. Its adoption for urban planning and policy crafting, however, is still lagging, mostly due to the highly interdisciplinary endeavor involved in understanding the role, and impact, of mobility across the social, technological and ecological fabric of urban life \cite{Entwisle2007,Shelton2015,Salganik2018}.

So far, different conclusions have emerged when describing the spatial context of social interactions and, while important strides have been made, explaining how urban demographics and socio-economic indicators relate to mobility still remains a challenge. Early work explicitly shows that existing correlations between mobile phone usage and wealth may be a starting point towards using Information and Communication Technology (ICT) data for planning were sensitive data is available for research \cite{Blumenstock2015, FriasMartinez2012}. When spatial context is explicitly considered, mobility research, produced from different disciplines, seem to indicate that diversity of human trajectories across the city is a conserved trait among social groups sharing similar status (i.e. social, economic, etc.) \cite{Lenormand2015,Pappalardo2016,Alessandretti2018,Cottineau2019,Lenormand2020a,Barbosa2020}, albeit important differences exist across gender \cite{Jiron2009,Lenormand2015,Gauvin2020}, income \cite{Lotero2016, Pappalardo2016}, residential location \cite{Shelton2015} and other aspects of human life \cite{Palmer2013}. 

What is now accepted, despite early predictions of a decline in the importance of space with the emergence of information and communications technologies in the sixties \cite{McLuhan1964,Yeung1998}, is that 'real' social interactions connecting and exchanging wisdom, goods and affection are highly relevant to explain the hierarchical patterns of mobility \cite{Alessandretti2020}. In fact, recent studies have shown the high predictive power of social ties to describe activities, interests and locations in ego networks \cite{Bagrow2019,Song2010}. Hence, the notion that functional relationships between social networks and space are (strongly) mediated by the spatial opportunities available for human interaction seem to prevail across the literature \cite{Urry2003,Netto2017}. We also know, that while social interactions are deeply associated to mobility, they only represent a limited fraction of movements across the city \cite{Cho2011} hinting towards the existence of other components associated to mobility and social mixing. It is also becoming clear that multivariate analyses of the factors linked to travel schedules, while important, often provides only localized descriptions hampering generalizations of the phenomena compared to ICT traces that explicitly measure how individuals use urban spaces during their daily journey \cite{LeRoux2017,Dannemann2018}. This has made ICT tracers great candidates to deepen our understanding of the dynamics of social mixing and the spatial environment in which they are embedded \cite{Carrasco2008, Gonzalez2008}.

We here study the socioeconomic structure of spatial and social interactions using mobile phone records of a major provider in Chile. We begin by extracting the spatial and social networks of interactions. We then introduce an indicator, akin to an urban pulse \cite{Miranda2017}, to assess the weekly mobility patterns of every urban locations in Chile. We use this indicator to cluster the locations showing similar weekly mobility patterns. We obtained four spatial clusters, strongly correlated with the socioeconomic status of its residents which finally allow us to build and analyze coarse grained spatial and social interaction matrices showing the emergence of a preferential association in terms of spatial and social interactions between people sharing similar socioeconomic status.

\begin{figure*}
	\centering 
	\includegraphics[width=\linewidth]{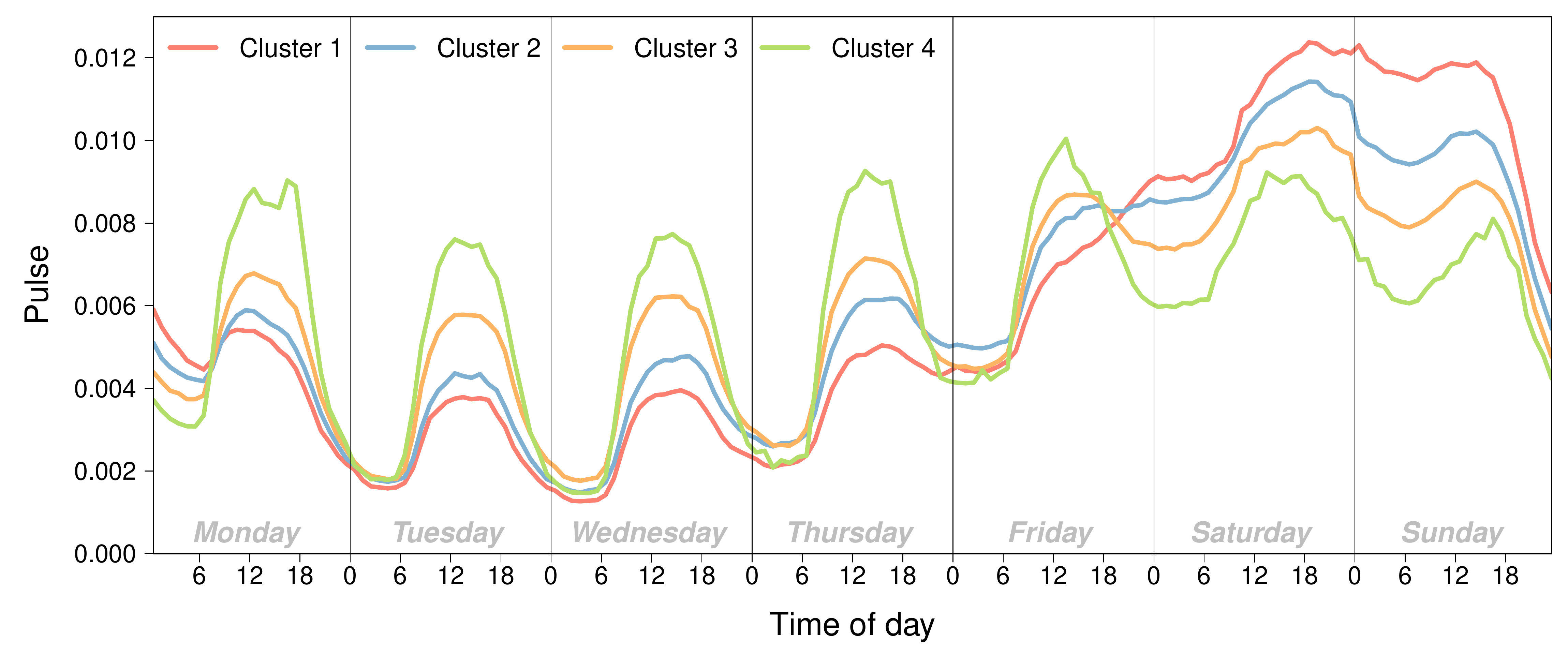}
	\caption{\textbf{Average pulse associated with the four clusters.} Plots displaying the standard deviations are available in Figures S7 and S8. It is worth noting that the fraction of reliable users (i.e. mobile phone users with a validated home location) is stable between the different clusters (Figure S9 in Appendix). \label{Fig1}}
\end{figure*}

\section*{Materials and Methods}

\subsection*{From data to networks}

Our datasets are composed of Data Detail Records (XDR) and Call Detail Records (CDR) provided by Telef{\'o}nica Chile representing 37\% share of the mobile phone market in Chile. The XDR dataset consists in billion of cellphone pings made by 4 million of mobile phones during 3 weeks in March, May and October 2015 in Chile. Each ping is characterized by its location (i.e. cellphone tower) and a timestamp. Each week has been divided into $T=168$ hours. We partitioned the country into $L=3,876$ locations following a Voronoi tessellation based on the cell phone towers' position. Data processing started by identifying the mobile phone users' home location for each week of observation \cite{Lenormand2016a}. We finally selected 2.5 million of reliable users with a \textit{validated} home location for at least one of the three weeks. We removed users for which home locations were not possible to identify. We were thus able to identify 360 million of spatial events defined as the presence of a reliable user in a location at time $t\in |[1,168]|$. This collection of spatial events has enabled us to build 168 spatial networks, one for each hour $t$. These networks are weighted and directed. A weight $G_{ij}^t$ of a link between two locations was given by the number of users living in location $i$ and that were present in location $j$ at time $t$ (i.e. all weeks combined). Similarly, we used the CDR dataset to identify 12.5 million of social events between reliable users. We defined a social event as a directed interaction (through a phone call) between two reliable users. In this case, we defined 168 social networks. The weight $S_{ij}^t$ of a link corresponds to the number of social interactions made by users living in location $i$ with users living in location $j$ at time $t$ for all weeks combined. More details regarding the data cleaning process are available in section Appendix (Table S1 and Figures S1 to S4).

\subsection*{Pulse of a location}

We characterize the weekly mobility pattern of a location with a spatio-temporal indicator that we called the 'Pulse of a location'. We define such pulse $P_i$ at location $i$ as the time-evolution of the average distance between the location $i$ and the position of its residence during a typical week. More specifically, the pulse $P_i^t$ of location $i$ at time $t\in |[1,168]|$ corresponds to the average distance between the location $i$ and the position of its residents at time $t$ (Equation \ref{pti}). 
\begin{equation}
	P_i^t=\frac{1}{A_i}\frac{1}{G_{i.}^t}\sum_{j=1}^{L} G_{ij}^t d_{ij}  
	\label{pti}
\end{equation}
where $L$ is the total number of locations, $d_{ij}$ the great circle distance between locations $i$ and $j$ and $G_{i.}^t=\sum_{j=1}^{L} G_{ij}^t$. The constant $A_i$ is used as normalization factor to ensure that $\sum_{t=1}^T P_i^t=1$. Note that a large heterogeneity of location areas exist, given irregular location of antennas across the study area. This prompted us to only consider pulses associated with the $2,294$ locations having a surface area lower than $10\mbox{ km}^2$ in order to compute pulses representative of the spatio-temporal status of the population.

\subsection*{Cluster analysis}

We rely on the ascending hierarchical clustering (AHC) method to identify different profiles of pulse across locations. Ward's metric and Euclidean distances were taken as agglomeration method and dissimilarity metric, respectively \cite{Hastie2009}. The number of clusters was chosen by comparing the ratio between the within-group variance and the total variance. The purpose of this cluster analysis is to identify meaningful profiles of pulse that can be used as a proxy for the socio-economic structure of a location. Indeed, we make the assumption that differences in mobility behaviors and particularly between week days and weekends represent an important descriptor of the socioeconomic status of the location.

\subsection*{Measuring spatial and social interactions}

We construct two coarse-grained spatial and social interaction matrices $\lambda$ and $\gamma$ based on the aggregation of link weights, $G_{ij}^t$ and $S_{ij}^t$, in space and time. More specifically, the fraction of spatial interaction from a cluster $c$ to a cluster $c'$ during a given time window $\Delta_t$ is defined as follow,
\begin{equation}
	\lambda_{cc'}=\frac{1}{B_c}  \sum_{\scriptstyle i \in c} \sum_{\scriptstyle j \in c' \atop \scriptstyle j \ne i} \sum_{\scriptstyle t \in \Delta_t} G_{ij}^t
	\label{lambda}
\end{equation}
where $\Delta_t$ is the set of hours contained in the time window. The constant $B_c$ is used as normalization factor to ensure that the sum of interactions from a cluster $c$ to the $N$ clusters is equal to one, $\sum_{c'=1}^{N} \lambda_{c,c'}=1$. The same formula is used to compute the social interactions $\gamma_{cc'}$ between and within clusters based on $S_{ij}^t$ instead of $G_{ij}^t$. 

To rigorously quantify the structure of these interactions, we use the index $\Phi$ proposed in \cite{Bassolas2019} to measure the hotspots' hierarchical structure of cities. In our case, this index allows to quantify the importance of interactions between close clusters (i.e. $|c-c'| \leq 1$) among all interactions as the index relies on the tridiagonal trace of the matrix $\lambda$ (Equation \ref{phi}, where $\delta_{cc'}$ is the Kronecker delta). Such approach provides a succinct representation of the preferential relationships between locations across the study area.  The same procedure is used to compute the index, now associated with the social interactions, by using $\gamma$ instead of $\lambda$ in the formula.
\begin{equation}
	\Phi=\frac{\sum_{c,c'=1}^{N} \lambda_{cc'} (\delta_{cc'} + \delta_{c(c'-1)} +     \delta_{(c-1)c'})}{\sum_{c,c'=1}^{N} \lambda_{cc'}}  
	\label{phi}
\end{equation}
The values of $\Phi$ range from 0 to 1. A value of 1 means that all the elements of the matrix that are not on the tridiagonal are equal to 0. In other words, all the interactions are occurring within the same cluster or with the closest cluster. A value of 0 means that the tridiagonal trace of the matrix is null, implying an abscence of interactions within the same cluster or with the closest cluster. However, this specific case is clearly  unrealistic so to rescale the value of $\Phi$ in a relevant order of magnitude we proposed the following  min-max normalization to obtain the metric $\bar{\Phi}$ (Equation \ref{phibar}).
\begin{equation}
	\bar{\Phi}=\frac{\Phi_h-\Phi}{\Phi_h-1} 
	\label{phibar}
\end{equation}
where $\Phi_h$ is the index obtained with a null model based on Equation \ref{lambda} in which a cluster is randomly assigned to every location (preserving the total number of locations per cluster). The value of $\Phi_h$ is then averaged over $100$ random reassignments. $\bar{\Phi}$ varies from 0, when the proximity between clusters is equivalent to the one obtained with the null model, to 1, when only interactions between nearby clusters occur. More details regarding the impact of the number of random reassignments used to compute $\Phi_h$ on $\bar{\Phi}$ are available in Figure S10 in section \emph{Null model} in Appendix. 

\begin{figure*}
	\centering 
	\includegraphics[width=\linewidth]{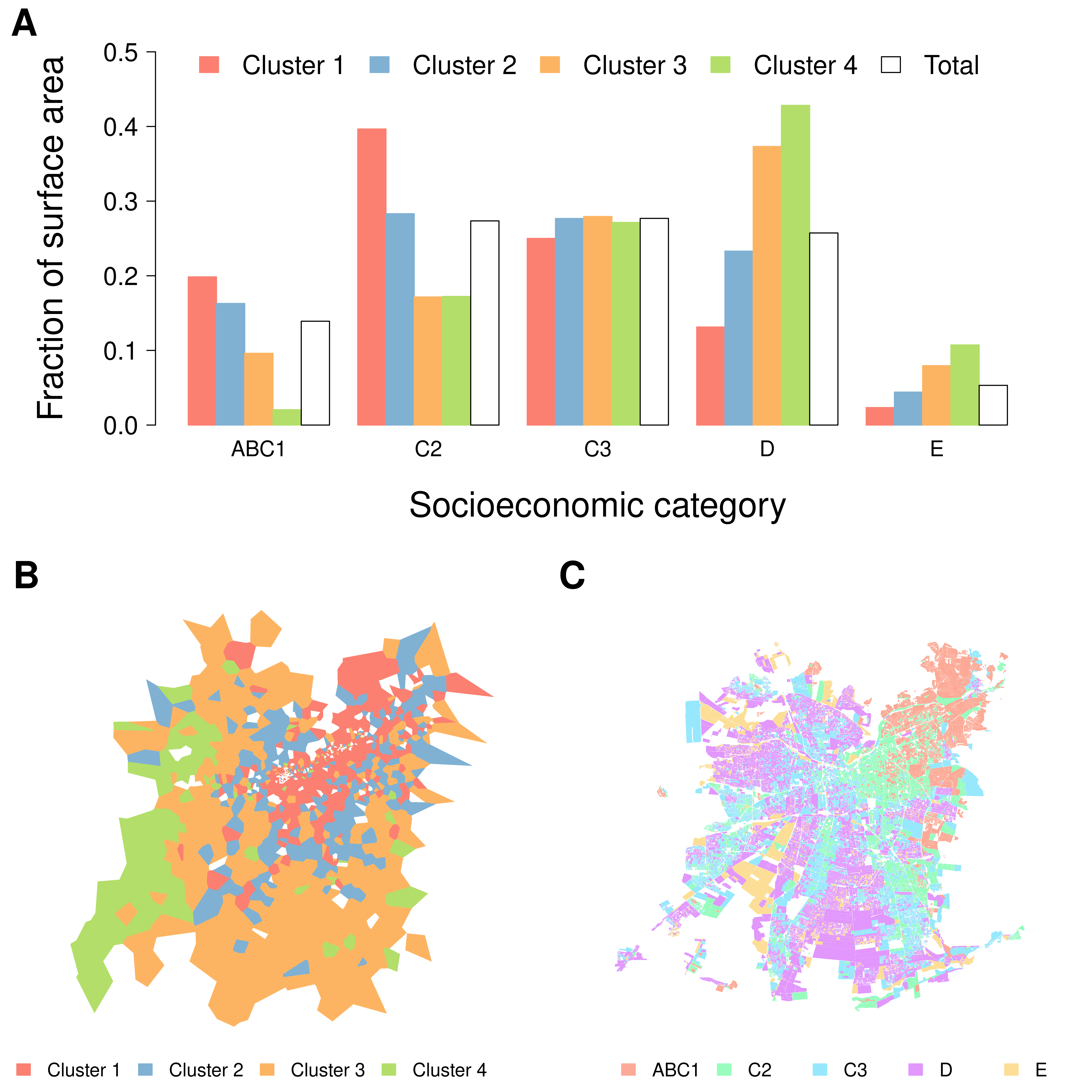}
	\caption{\textbf{\textbf{Socioeconomic characteristic of the clusters.}} (\textbf{A}) Fraction of surface area dedicated to each socioeconomic category according to the cluster (colored bars) and in total (white bar) for the whole country. (\textbf{B}) Maps of the four clusters in Gran Santiago (the largest city). (\textbf{C}) Spatial distribution of socioeconomic categories in Gran Santiago (the largest city). \label{Fig2}}
\end{figure*}

\section*{Results}

\subsection*{Pulse of a location and socioeconomic structure}

Based on the ratio between the within-group variance and the total variance (Figure S5 in Appendix), 18 clusters were found. As can be seen in Figure S6 in Appendix, $92$ percent of the location are covered by four main clusters. The four average pulses associated with these clusters are displayed in Figure S7 in Appendix. The rest of the locations are gathered into 3 small clusters (Figure S8 in Appendix) and 11 outliers (category \textit{Others} in Figure S6) that we decided to discard because they contain too few locations (or even one location for the outliers) to allow for a rigorous analysis.

Thus, we obtain four main pulse' profiles gathering 92\% of the locations. Figure \ref{Fig1} shows a profile of the average pulse activity for each of these four clusters. Not surprisingly, each profile exhibits a typical day-night temporal activity pattern where individuals are moving, on average, further away from their residence during the day compared to night hours. Some differences can nevertheless be observed between the different days of the week. The average distance from home tend to increase from Wednesday to Saturday and then decrease from Sunday to Tuesday. The difference between day and night is also more pronounced on week days compared to weekends. The main difference between profiles is mostly based on the difference in mobility behaviors between week days and weekends. This difference is very pronounced for the locations belonging to the cluster 1 (representing 25\% of the locations). Indeed, people living in locations belonging to cluster 1 tend to roam farther away from home during weekends compared to week days. This difference is slightly decreasing for the 26\% and 34\% locations belonging to cluster 2 and 3, respectively. The opposite behavior is observed for people living in cluster 4 (7\% of locations) that tend to be more or less at the same distance from their home irrespective of the day of the week. This pattern is congruent with descriptions of individual mobility journeys in which working class groups tend to exhibit longer journeys to work compared with more affluent sectors in Chile \cite{Jiron2009,Jiron2020}. 

\begin{figure*}
	\centering 
	\includegraphics[width=\linewidth]{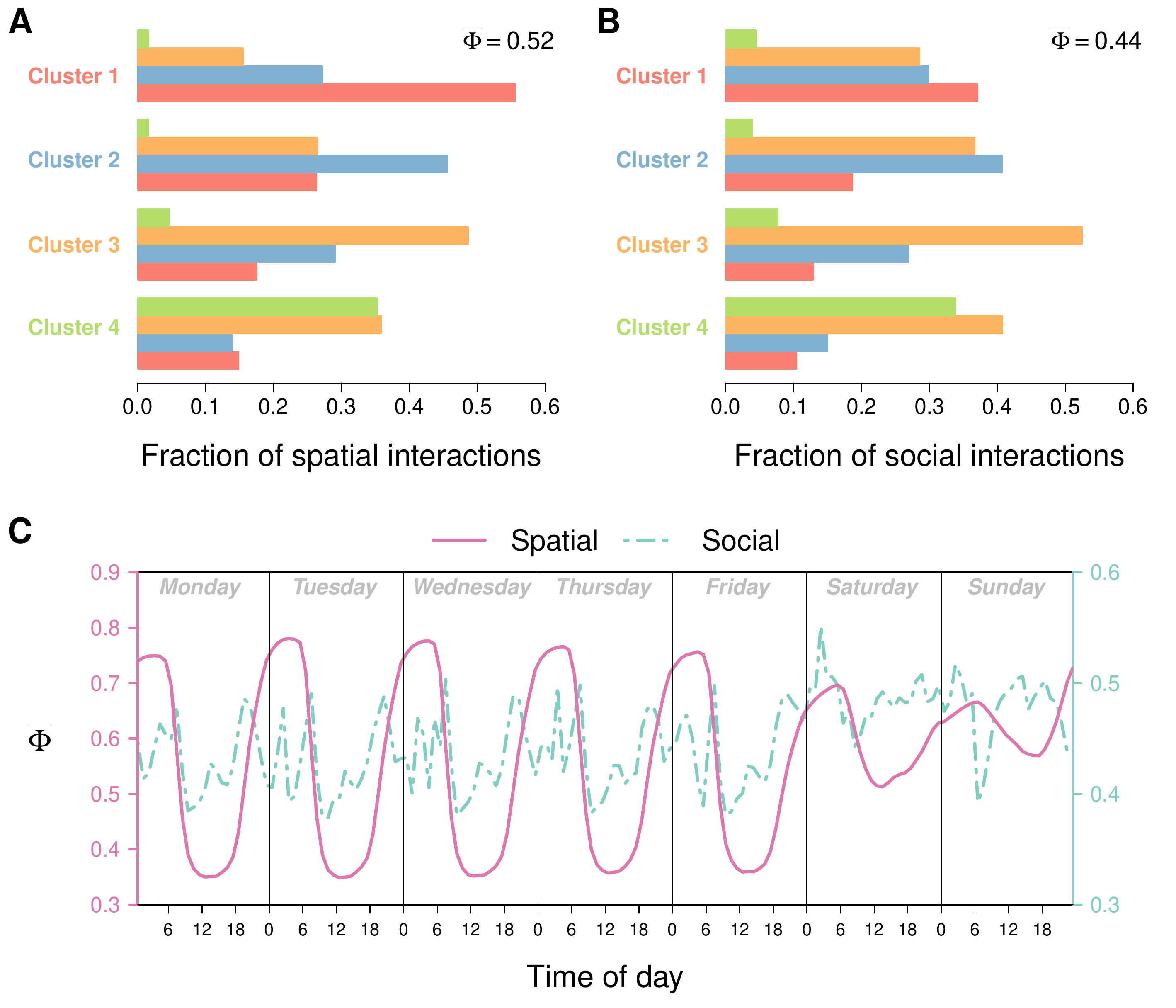}
	\caption{\textbf{Socio-spatial interactions analysis.} (\textbf{A}-\textbf{B}) The fraction of spatial (\textbf{A}) and social (\textbf{B}) interaction within and between clusters. The value of $\bar{\Phi}$ obtained with both matrices are displayed. (\textbf{C}) Temporal evolution of $\bar{\Phi}$ across week hours for the spatial interactions (in pink) and social interactions (in green). \label{Fig3}}
\end{figure*}

In order to understand the origin of the observed differences in mobility behavior between week days and weekends, we investigated the relationship between the pulse of a location and its socioeconomic status. To do so, we attach to each location the socioeconomic structure of its residents (when the information was available). The indicator used is divided into five relevant socioeconomic categories labeled ABC1, C2, C3, D and E with ABC1 as the most wealthy group and E the group with the lowest income and educational level. The socioeconomic structure of a location is based on the surface area dedicated to the socioeconomic category of each census track intersecting the location (see section \emph{Socioeconomic structure of the locations} in Appendix for further details). The relationship between these four clusters and the five socioeconomic categories is plotted in Figure \ref{Fig2}. We observe in Figure \ref{Fig2}A how the fraction of surface area of locations belonging to a given cluster is distributed among the socioeconomic categories for the whole country. It is worth noting that a socioeconomic gradient exists from cluster 1, characterized by an over-representation of wealthy neighborhoods (i.e. comparatively larger red bars for the categories ABC1 and C2), to cluster 4 which shows an over-representation of neighborhoods with low incomes and educational level (i.e. larger green bars for categories D and E). The comparison of the spatial distribution of clusters (Figure \ref{Fig2}B) and socioeconomic categories (Figure \ref{Fig2}C) in Gran Santiago (the largest city) confirms these results. Indeed, this particular spatial pattern of socioeconomic distribution has been described with details in the literature, with a concentration of more affluent neighborhoods projected in a cone-shaped area that starts at center of the city and opens towards the east and northeast outskirts of Santiago. This particular spatial segregation pattern has recently been corroborated by newer research \citep{Dannemann2018,Garreton2020}. This pattern is particularly apparent while looking at the spatial distribution of cluster 1 and 2 in Figure \ref{Fig2}B.

\begin{figure*}
	\centering 
	\includegraphics[width=\linewidth]{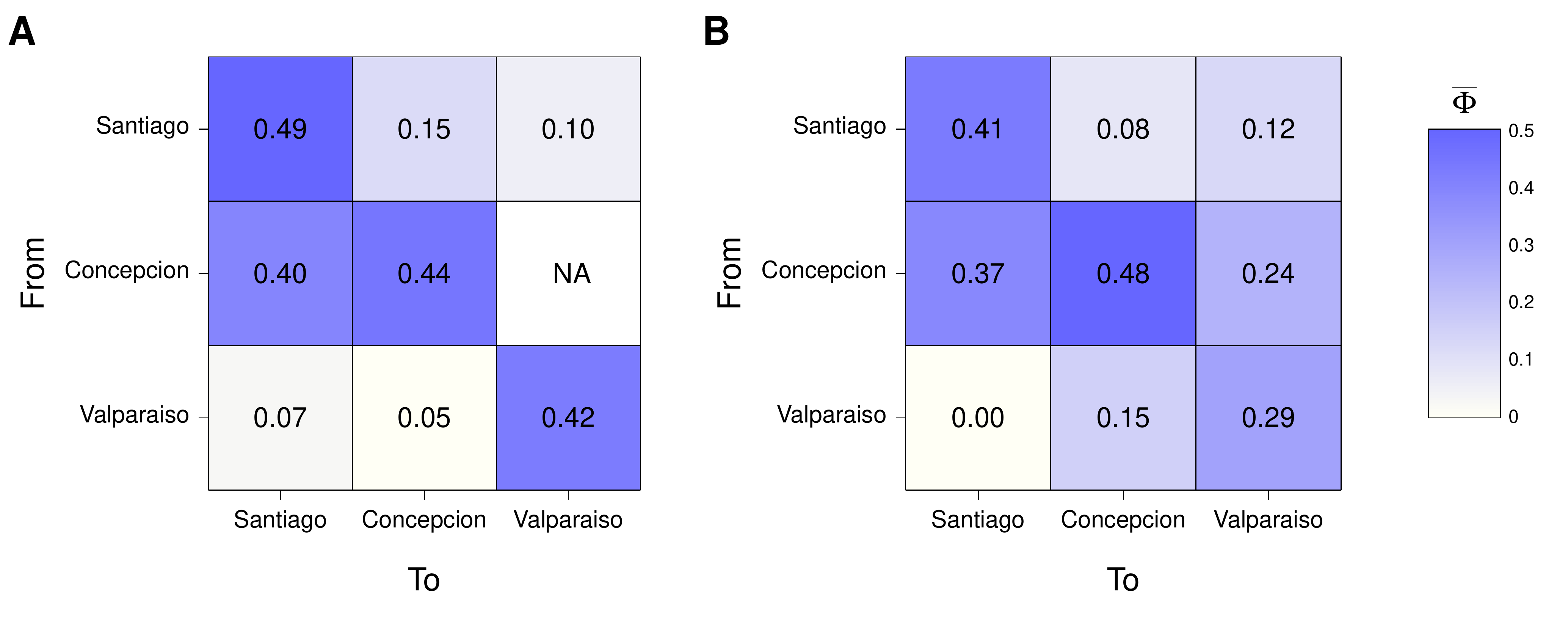}
	\caption{\textbf{Intra-city and inter-cities socio-spatial interactions analysis.} The index value are based on spatial (\textbf{A}) and social (\textbf{B}) interactions between locations in the same city (diagonal) or from one city to another. There were not enough data available (NA) to measure the spatial interactions between Concepci\'on and  Valpara\'iso. \label{Fig4}}
\end{figure*} 

\subsection*{Socio-spatial interactions analysis}

The results obtained for a week window period (i.e. $\Delta_t=|[1,168]|$) are presented in Figure \ref{Fig3}A and \ref{Fig3}B. Each bar represents an element of the interaction matrices and can be interpreted as the probability of spatial and social interactions between two clusters during a typical week. The figures indicate that locations belonging to the same cluster -- or similar clusters -- tend to mostly interact with each other compared to their interaction with other locations, both spatially and socially. We also observe that these preferential interactions are less marked for social interactions (Figure \ref{Fig3}B) than for the spatial ones (Figure \ref{Fig3}B). 

Our results show a $\bar{\Phi}$ value of $0.52$ for the spatial interaction matrix (Figure \ref{Fig3}A) and $0.44$ for the social interaction matrices (Figure \ref{Fig3}B). These values demonstrate that a clear proximity exists in terms of spatial and social interactions between locations sharing similar socioeconomic features. It also shows that such pattern is not just driven by spatial constraints. In other words, these results clearly show that people living in locations of a given socioeconomic status tend to move in, and socially interact with, people living in locations of the same, or similar, socioeconomic status. While slightly higher for the spatial interactions than for the social ones, it is particularly remarkable that both $\bar{\Phi}$ values are quite high, as this metric intrinsically considers a random model of interactions that effectively considers spatial autocorrelation. That is, it explicitly considers what could happen in a random situation.

In order to deepen the socio-spatial interactions analysis, we plot in Figure \ref{Fig3}C the temporal evolution of $\bar{\Phi}$ during a typical week using a time window of one hour (see Equation \ref{lambda}). As expected, the value of $\bar{\Phi}$ varies greatly according to the day of the week and the hour of the day. A greater variations is observed for the spatial interactions compared to the social ones. During weekdays, the spatial proximity between cluster is higher during the night with a $\bar{\Phi}$ value going from 0.75-0.8 compared to the 0.35 observed during the 11-19 hour span. 
The variations decrease during weekend days with less proximity during night hours ($\bar{\Phi}=0.7$) and more during the day ($\bar{\Phi}=0.6$). This result also suggests that structural dependence between clusters, revealed by $\bar{\Phi}$, is more relevant when everybody is at home, confined to their individual socio-economic groups.
Social interactions, in turn, show more nuanced and noisy results. Presumably given the comparatively lower number of social events during the night (see Figure S4). Nevertheless, we also observe that $\bar{\Phi}$ value decrease during weekdays. In such time span, particularly during the morning, people tend to interact less with people living in a similar cluster, a pattern that increases during the evening hours. However, it is interesting to note that this increase of social interactions with people living in a similar cluster starts earlier over the course of the day. It is also characterized by two peaks, one halfway through the day and another one around 6pm. During weekend days, $\bar{\Phi}$ is quite stable with a value fluctuating around 0.5.  

Finally, Figure \ref{Fig4} shows the $\bar{\Phi}$ index for intra- and inter-city interactions. In this case, an additional constrains is added in the Equation \ref{lambda} to only consider interactions between locations that belongs to cluster $c$ in one city with locations belonging to cluster $c'$ in the same or in another city, hence highlighting same-cluster interaction. We focus here on the three largest cities in terms of population in Chile. As it can be observed in Figure \ref{Fig4}, the value of $\bar{\Phi}$ capturing spatial (Figure \ref{Fig4}A) and social interactions (Figure \ref{Fig4}B) between locations in the same city are in line with $\bar\Phi$ values obtained for the whole country. We also note that these preferential spatial and social interactions hold for several pairs of cities such as the people living in Concepci\'on interacting with locations and people living in Santiago.

\section*{Discussion}

This study not only concurs to other studies showing how mobile phone data may aid to shape a better understanding of the socioeconomic structure of spatial and social interactions in urban system, but it also proposes a methodological approach to assess the hierarchical structure of spatiotemporal interactions across the city. By defining two temporal networks representing interactions stemming from highly resolved spatial and social events, we are able to describe how people ascribed to a particular socioeconomic levels within the city interact with their environment and with people living in other locations across weekly hours. Similarly to \cite{Wang2018}, the net result here shows that people living in locations of a given socioeconomic status preferentially interact with locations and people sharing the similar socioeconomic levels. Additionally, while this proximity varies similarly for both spatial and social interactions during the course of the week, social interactions measured by the voice calls between users exhibit a more nuanced association between socioeconomic status, much like what has recently been described in the literature \cite{Xu2021}. This may be the product of a combination of factors, including the fact that the events captured by our voice call dataset is composed by a combination of professional, personal and leisure interactions that may increase social mixing. 

Our study shed new lights on the understanding of social mixing using large datasets. In fact, the mounting availability of such type of information is contributing to make large strides to describe the effect of segregation on the various realms of our society \cite{Mena2021, Gauvin2020, Beiro2018, Lenormand2020a}. While our results contribute to the analysis and understanding of the relationship between urban mobility, social networks and socioeconomic status, they also raise a number of new questions with regard to their generalization. In this regard, we will argue that the large sample used in this analysis, in terms of spatial and social events attached to a substantial number of mobile phones users in Chile (See Table S1), provides an empirical description on how socioeconomic status relates to the spatial and social interactions at several levels. 

For instance, the role of space have been a central topic in the understanding social tie formation. At local scale, space has been described to determine interactions through distance, urban configuration \cite{Wang2018a,Liu2022} and specific locations fostering social interactions \cite{Small2019}. These studies highlight not only the importance of space-mediated interactions, but also the relevance of social interactions such as relationship maintenance among friends \cite{Carrasco2008}. While these conclusions go beyond this particular work, we envisage that ongoing improvements in the identification of residences \cite{Pappalardo2021} and transportation modes \cite{Graells2018}, among others, will clearly foster more granular descriptions of urban dynamics. In fact, they may even shift the focus to more localized descriptions of social and transportation behaviour, as has recently been seen by the analysis of the ongoing COVID-19 pandemia in Chile \cite{Mena2021, Gozzi2021}. At broader scales, spatial limitations (e.g. Modifiable Areal Unit Problem) have recently been invoked to highlight the difficulties describing spatial aspects of segregation \cite{Garreton2020}. While this old geographic issue may certainly hamper the possibility to inform social mixing from mobile phone datasets, other equally important aspects of areal distributions may acquire relevance in the spatially explicit descriptions of cities. For instance, the definition of urban entities may also concur to MAUP to describe the correct functional extension to which urban descriptions should be attached    \cite{Cottineau2019,Sotomayor2020}. In spite of this, it is interesting to note that the preferential interactions among socioeconomic status in Chile, as described here, appear to be holding even when considering interactions between cities hinting towards an intrinsic property of social systems as opposed to a particular constraints (e.g. spatial) imposed on the interaction network \cite{Onnela2011}. 

Finally, it is also worth noting that the usage of the hierarchy index proposed in \cite{Bassolas2019} used here, provides a simple conceptual mean to compare both, social and spatial, networks across the whole country that is independent of urban shape, while still capturing the spatial hierarchy of mobility within and between cities \cite{Pumain2018}.

\section*{Acknowledgments}

The work of ML was supported by a grant from the French National Research Agency (project NetCost, ANR-17-CE03-0003 grant). HS was supported by the Chilean Agency of Research and Development ANID (FONDECYT Regular grant \#1211490). Thanks to Isidro Puig from the OCUC for his help on census data.

\bibliographystyle{unsrt}
\bibliography{Pulse}

\onecolumngrid

\makeatletter
\renewcommand{\fnum@figure}{\sf\textbf{\figurename~\textbf{S}\textbf{\thefigure}}}
\renewcommand{\fnum@table}{\sf\textbf{\tablename~\textbf{S}\textbf{\thetable}}}
\makeatother

\setcounter{figure}{0}
\setcounter{table}{0}
\setcounter{equation}{0}

\newpage
\clearpage
\newpage
\section*{Appendix}

\subsection*{Call and Location History} 

The data used in this study consists in Call Detail Records (CDR) and Data Detail Records (XDR) provided by Telef{\'o}nica Chile representing 37\% share of the mobile phone market in Chile. 

Our first dataset is composed of billions of cellphone pings made by 4 million of mobile phones during $3$ weeks in March, May and October 2015 in Chile. Each ping is characterized by its location (i.e. voronoi cell) and a timestamp informing us on the hour, the day of the week and the week when the ping has occurred. We structured the dataset in a four-column Location History Table (Week, Hour, User, Location). Each line represents a \textit{spatial event} informing us on the presence of a user in a location during a given week at a given time. If the presence of a user was detected into several location during the same hour, we chose the location with the highest number of events. In the event of a tie, one of them was drawn at random. 

In addition, we relied on a second dataset to compute the history of calls between mobile phone users. This dataset was structured into a four-column Call History Table (Week, Hour, Caller, Callee). Each line represents a \textit{social event}. A social event is characterized by a phone call made by a caller to a callee during a given hour and a given week. This means that if a caller called several times the same callee during a given hour, only one social event has been considered.

\subsection*{Identification of the users' place of residence}

The first step consisted in identifying the users's place of residence to filter out users with a low number of spatial events and/or exhibiting irregular mobility patterns. For each of the three weeks periods and for each user, we applied the following procedure to extract the home locations:

\begin{itemize}
	\item First, we focused on the user's spatial events occurring during nighttime hours (between 9pm and 8am included). Only days of the week from Monday to Thursday were considered ($N=48$ hours in total). We note $N_u$ the number of events occurring during nighttime hours.
	\item We applied here a \textbf{first filter} by considering only users with a number of spatial events higher than a fraction $\delta_A=N_u / N$ of the total number of nighttime hours. 	
	\item We identified the location in which the user has been localized the highest number of spatial events during nighttime hours. We define this location as her or his home location. 
	\item A \textbf{second filter} was also implemented to select only users whose fraction of events occurring at their home location during nighttime is larger than a fraction $\delta_R$ of the total number of events during nighttime.
\end{itemize}

As explained in \cite{Lenormand2016a}, the first filter $\delta_A$ is applied to discard users having a too low number of spatial events. The last filter allowed us to adjust the degree of confidence in the identification of the home location. We chose to fix $\delta_A$ to 0.3 and $\delta_R$ to 0.3 which seems to be a good interplay, allowing us to remove users not active enough and/or exhibiting irregular mobility patterns during the time period (Figure S1), while preserving the spatial distribution of inhabitants observed in Chile (Figure S2). The number of reliable users (i.e. with a validated home location) is available in Table S1.

\begin{figure}[!h]
	\centering 
	\includegraphics[width=\linewidth]{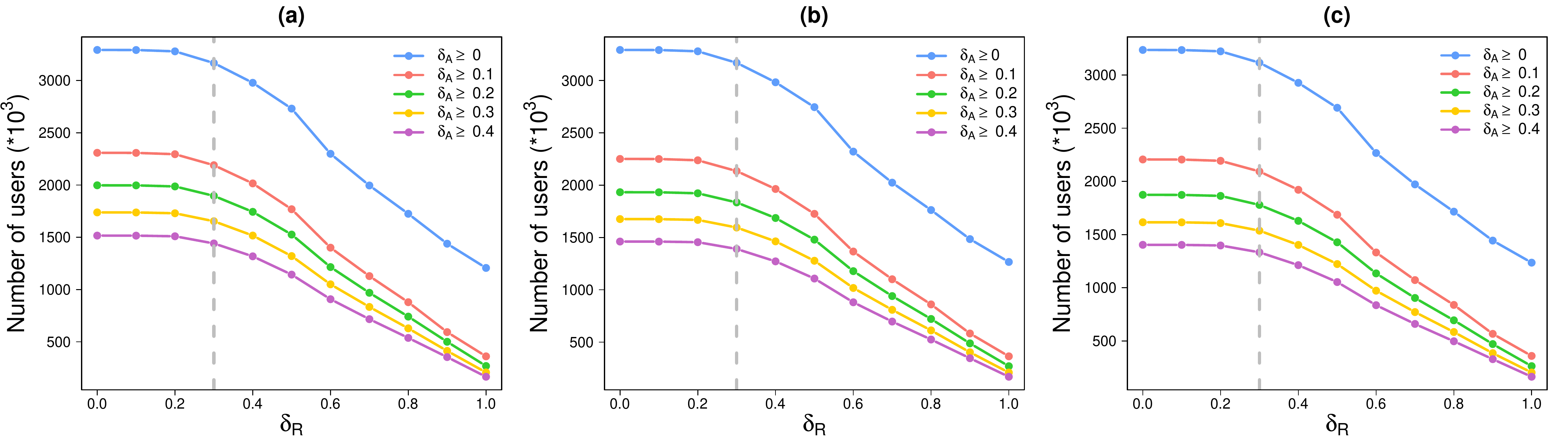}
	\caption{\textbf{Influence of the parameters.} Number of reliable users during the first (a), second (b) and third (c) week as a function of $\delta_R$ and for different values of $\delta_A$. The vertical bars indicate the value $\delta_R=0.3$. \label{FigA1}}
\end{figure}

\begin{figure}[!h]
	\centering 
	\includegraphics[width=\linewidth]{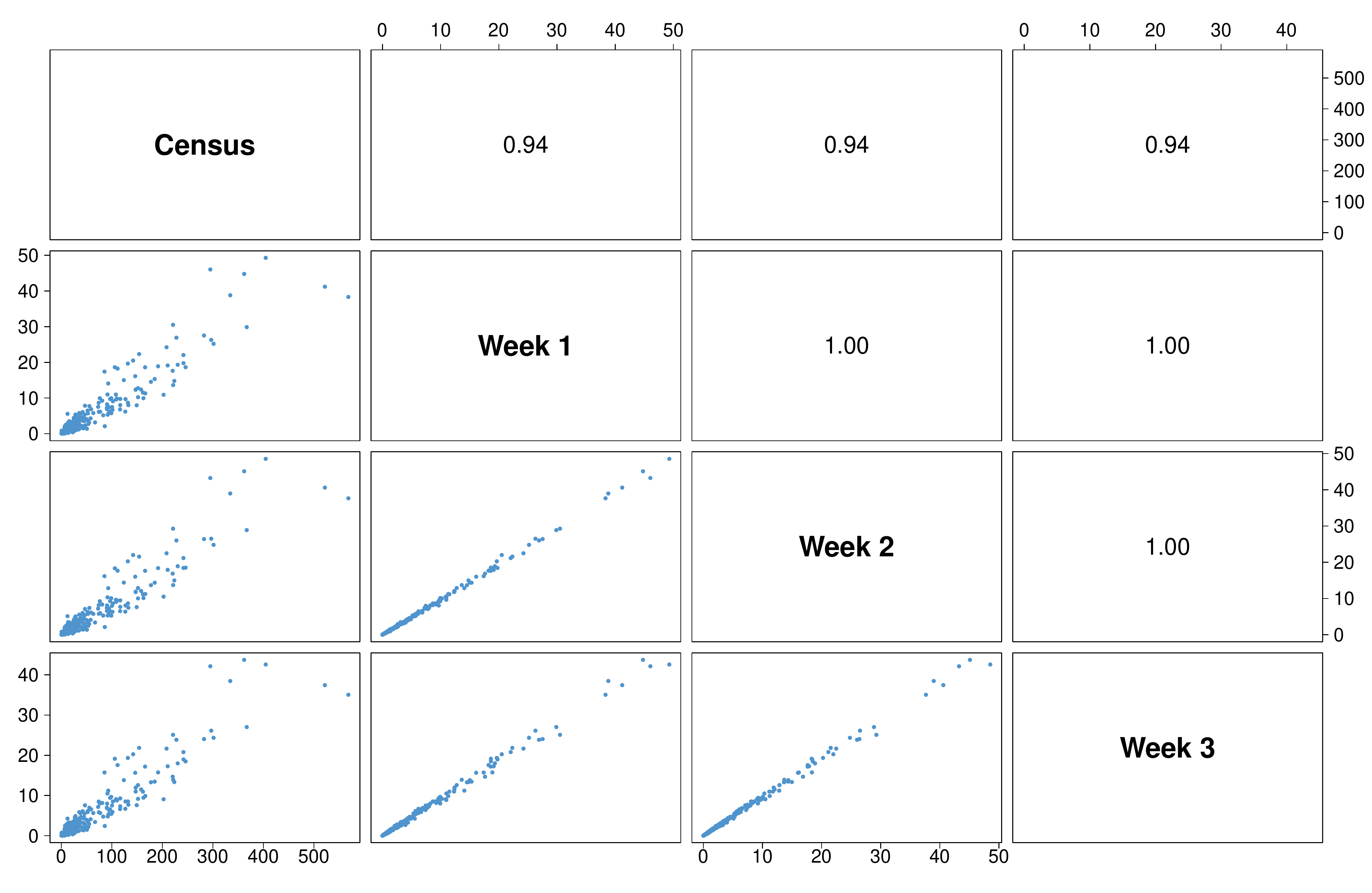}
	\caption{\textbf{Comparison between census and XDR data.} Each
	scatter plot and its associated Pearson correlation coefficient represents a comparison between the number of inhabitants (expressed in thousands of individuals) in the census and the number of inhabitants (expressed in thousands of individuals) estimated with XDR data (i.e. reliable users) during the three weeks of observation. Each point represents one municipality of Chile. \label{FigA2}}
\end{figure}

\begin{table}[!h]
	\caption{\textbf{Number of users (all) and reliable users according to the week of observation and in total.}}
	\label{TabA1}
	\begin{center}
		\begin{tabular}{lcc}
			\hline
			Date & \# Users (all) & \#Reliable Users \\
			\hline
			15 to 21 March 2015 & 3,292,923 & 1,657,048 \\
			10 to 16 May 2015    & 3,292,647 & 1,598,571 \\
			2 to 8 August 2015   & 3,236,122 & 1,539,621 \\
			\hline
			Total & 4,064,476 & 2,565,365 \\
			\hline
		\end{tabular}
	\end{center}
\end{table}

As mentioned in the previous section, there are some holes in the user history location with hours with no events. Nevertheless, we observe in Figure S3 that, during each week of observation, 75\% of the reliable users have at least 100 spatial events (60\% of the maximum value). 

\begin{figure}[!h]
	\centering 
	\includegraphics[width=12cm]{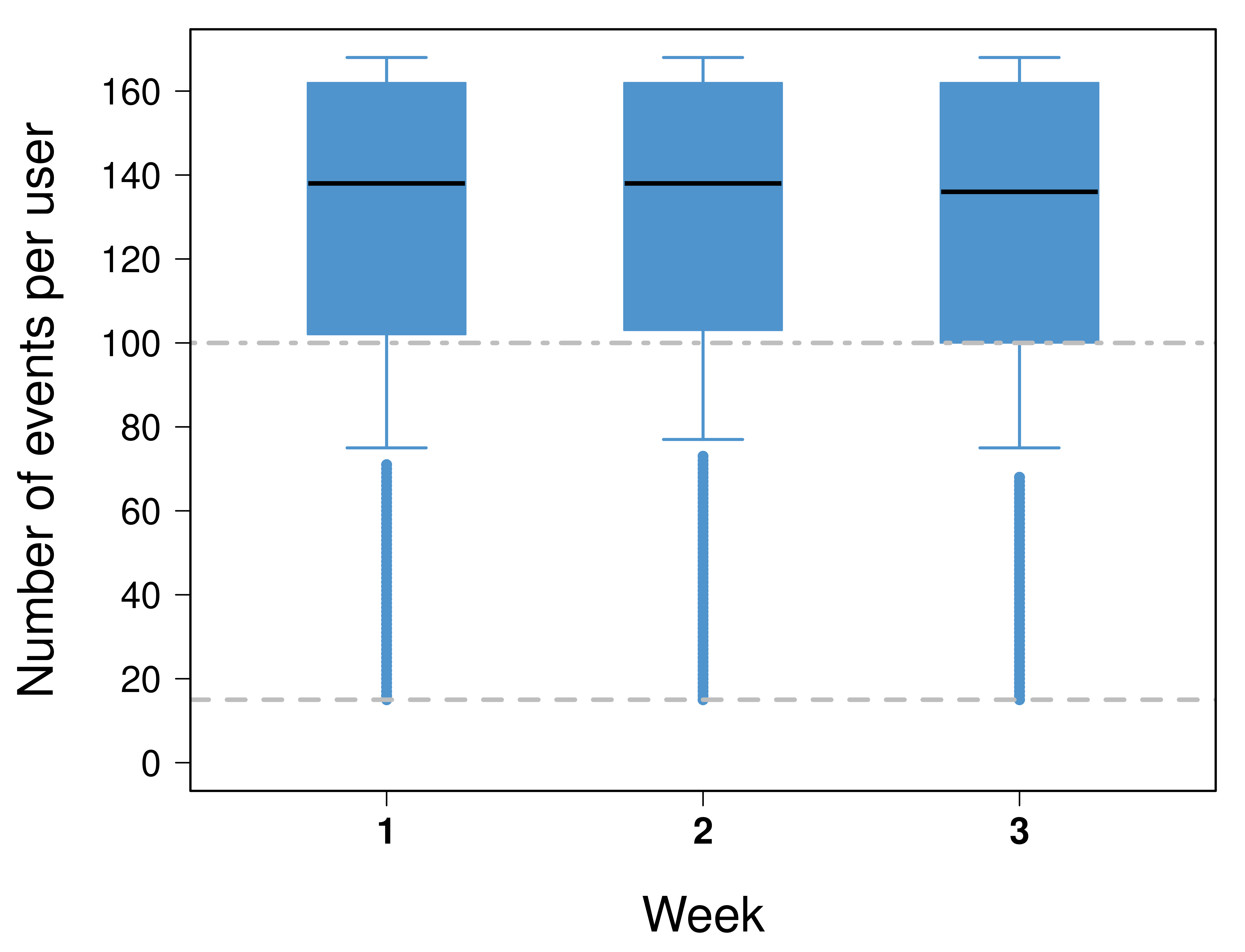}
	\caption{\textbf{Boxplots of the number of events per reliable user according to the week of observation.} The dashed grey line represents the minimum value (15 is the minimal value required to pass the first filter in the home identification). The dotdashed line represents the limit of 100 events. The maximum value is 168 (number of hours in the week). Each boxplot is composed of the first decile, the lower hinge, the median, the upper hinge and the last decile. The blue dots represents the outliers. \label{FigA3}}
\end{figure}

\clearpage
\subsection*{From events to networks}

A table summarizing the number of reliable users and their associated numbers of spatial and social events is available in Table S2. The associated temporal evolution is available in Figure S4. Finally, the collections of spatial and social events has enabled us to constructs $168$ spatial networks and $168$ social networks. The weight $G_{ij}^t$ of a link between two locations $i$ and $j$ at time $t$ is equal to the number of users living in location $i$ that were present in location $j$ at time $t$ (all weeks combined). Similarly, the link weight $S_{ij}^t$ of a social network is equal to the number of social interactions made by users living in location $i$ with users living in location $j$ at time $t$ (all weeks combined).

\begin{table}[!h]
	\caption{\textbf{Number of reliable users, spatial and social events per week and in total.}}
	\label{TabA2}
	\begin{center}
		\begin{tabular}{lccc}
			\hline
			Date & \#Reliable Users & \#spatial events & \#Social events \\
			\hline
			15 to 21 March 2015 & 1,657,048 & 129,760,887 & 4,433,505 \\
			10 to 16 May 2015   & 1,598,571 & 126,359,359 & 4,207,538 \\
			2 to 8 August 2015  & 1,539,621 & 120,960,807 & 3,905,935\\
			\hline
			Total  & 3,023,946 & 377,081,053 & 12,546,978 \\
			\hline
		\end{tabular}
	\end{center}
\end{table}

\begin{figure}[!h]
	\centering 
	\includegraphics[width=\linewidth]{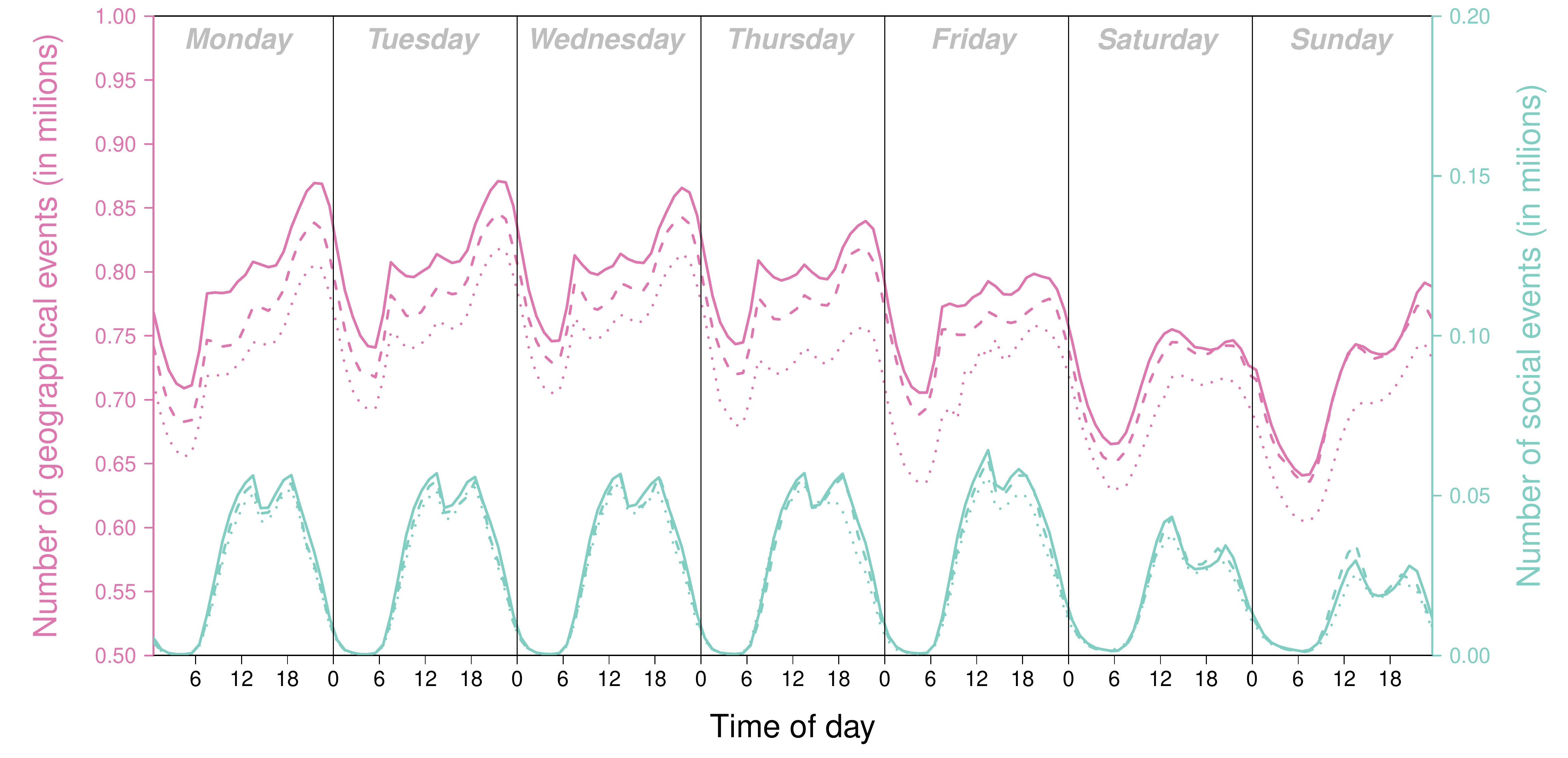}
	\caption{\textbf{Number of spatial events (in pink) and social events (in green) according to the hour of the day.} Each line represents a week of observation. \label{FigA4}}
\end{figure}

\clearpage
\subsection*{Socioeconomic structure of the locations}

As mentioned in the main text, we attached to each of the $3,876$ locations some information regarding the socioeconomic level of their residents when the information was available. To do so, we relied on the socioeconomic map of Chile proposed by Adimark \cite{Adimark2009,Dannemann2018}. These maps are available from the \textit{Observatorio de Ciudades UC} (OCUC) website (\url{https://ideocuc-ocuc.hub.arcgis.com/}, last accessed 06/12/2022 in Shapefile format for five major chilean cities.

\begin{itemize}
	\item Antofagasta in 2002 available at \url{https://ideocuc-ocuc.hub.arcgis.com/datasets/fbde68b6c3d547c8adfcc17d196e1e88_0}, last accessed 06/12/2022.	
	\item Coquimbo y La Serena in 2002 available at \url{https://ideocuc-ocuc.hub.arcgis.com/}, last accessed 07/01/20201. 
	\item Gran Concepci{\'o}n in 2002 available at \url{https://ideocuc-ocuc.hub.arcgis.com/datasets/f62f12fae97548fd8c71cb405d40e5f2_0}, last accessed 06/12/2022. 
	\item Gran Santiago in 2012 available at \url{https://ideocuc-ocuc.hub.arcgis.com/datasets/c264bc8bca7f45bc8ae74329557628b2_0}, last accessed 06/12/2022. 
	\item Puerto Montt and Puerto Varas in 2002 available at \url{https://ideocuc-ocuc.hub.arcgis.com/datasets/91deae3707ff447f961b4e2a5cf2300d_0}, last accessed 06/12/2022. 
	\item Valpara{\'i}so in 2002 available at \url{https://ideocuc-ocuc.hub.arcgis.com/datasets/b9458dbbc94343e58ea5fc9c5def03f9_0}, last accessed 06/12/2022.   				
\end{itemize}

These data inform us on the dominant socioeconomic categories of the resident of each 'manzana' (i.e. census block). There are five categories labeled ABC1, C2, C3, D and E with ABC1 as the most wealthy group and E the group with the lowest income and educational level. For each location we computed the area of the intersection between the Voronoi cell and the census blocks (if any) for each category. To identify the socioeconomic structure of each cluster, we computed the fraction of surface area (of the locations composing this cluster) dedicated to each socioeconomic category.

\clearpage
\subsection*{Clustering}

\begin{figure}[!h]
	\centering 
	\includegraphics[width=11cm]{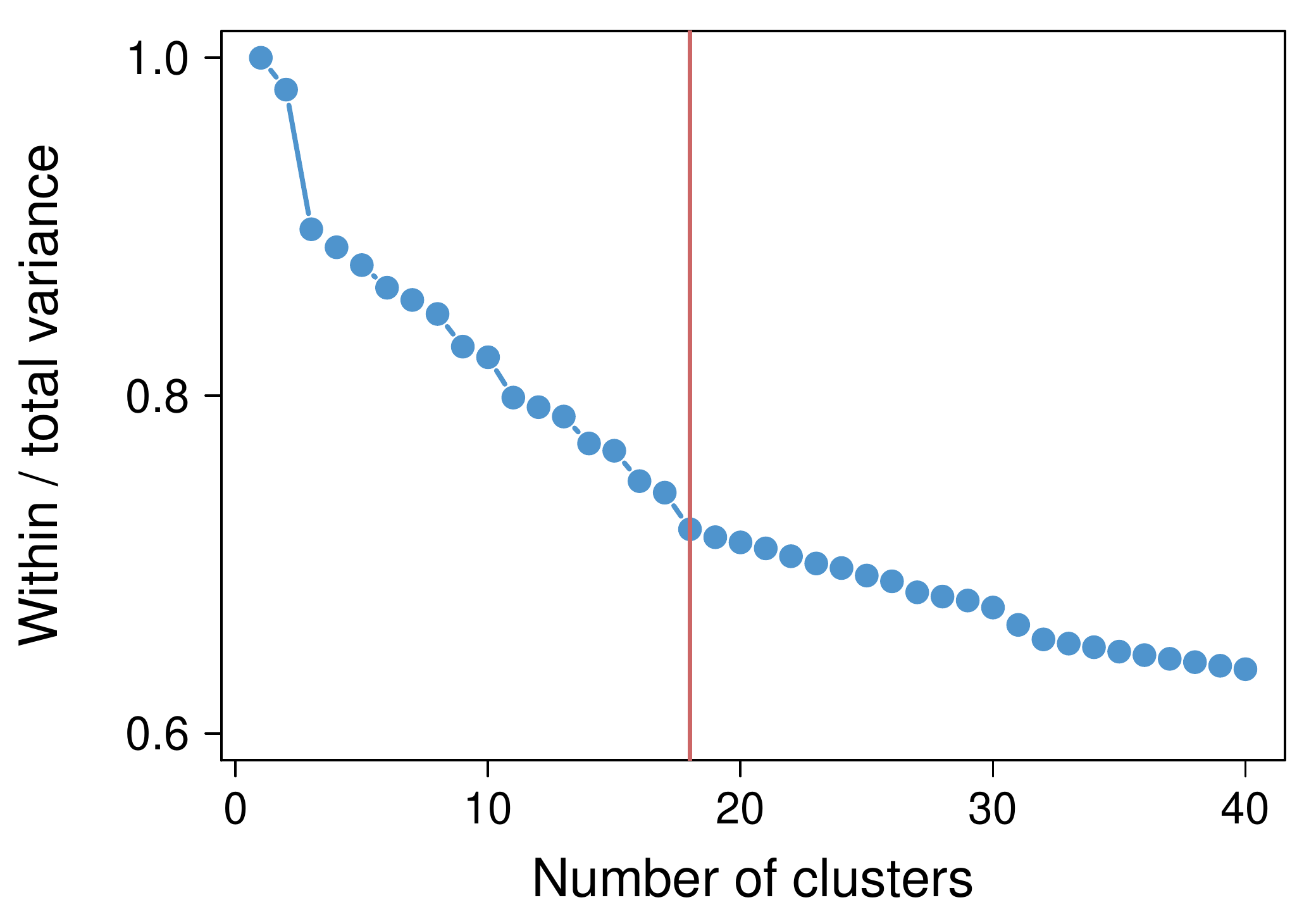}
	\caption{\textbf{Ratio between the within-group variance and the total variance as a function of the number of clusters.} \label{FigA5}}
\end{figure}

\begin{figure}[!h]
	\centering 
	\includegraphics[width=\linewidth]{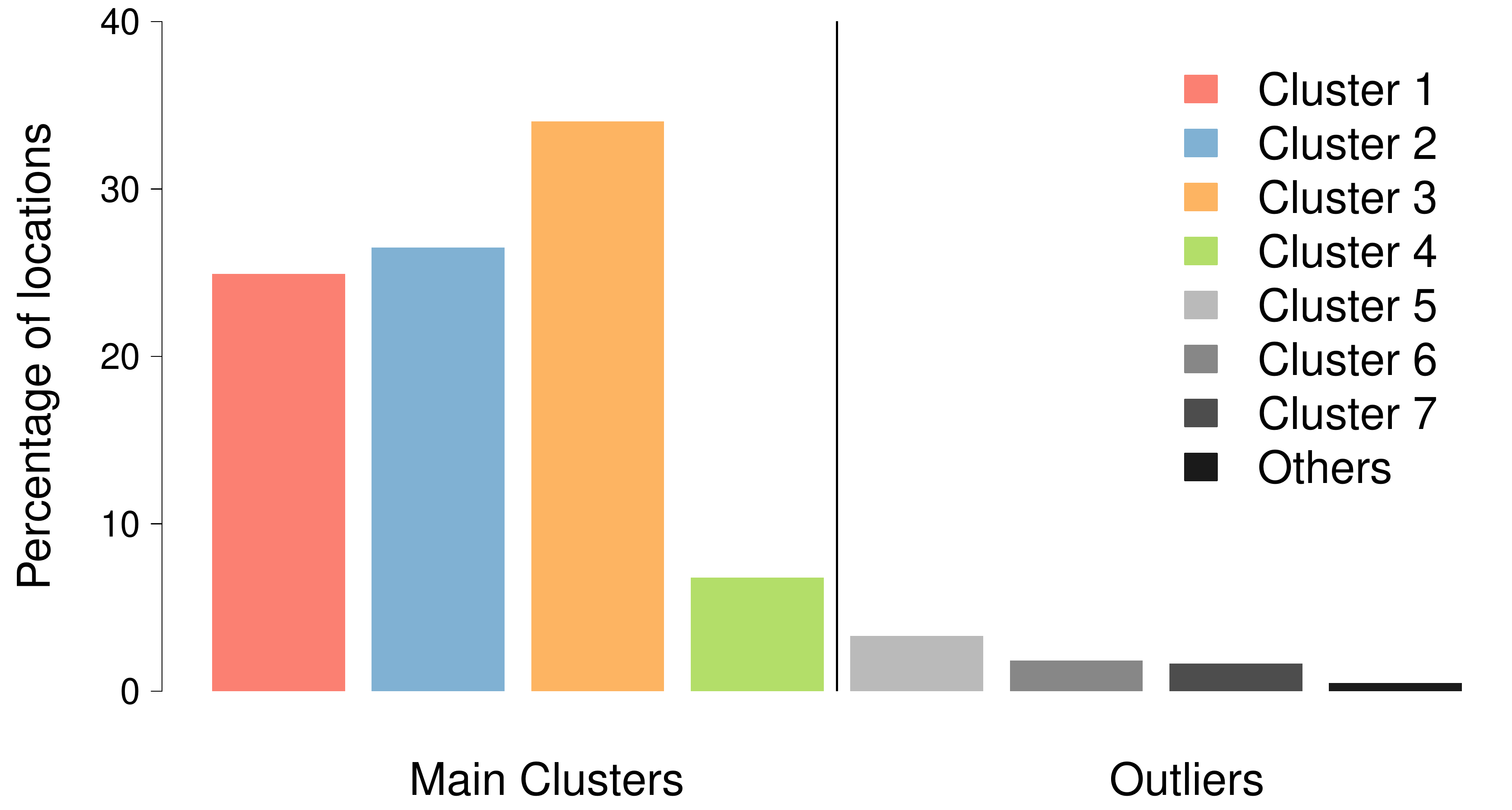}
	\caption{\textbf{Percentage of locations by cluster.} \label{FigA6}}
\end{figure}

\begin{figure}[!h]
	\centering 
	\includegraphics[width=\linewidth]{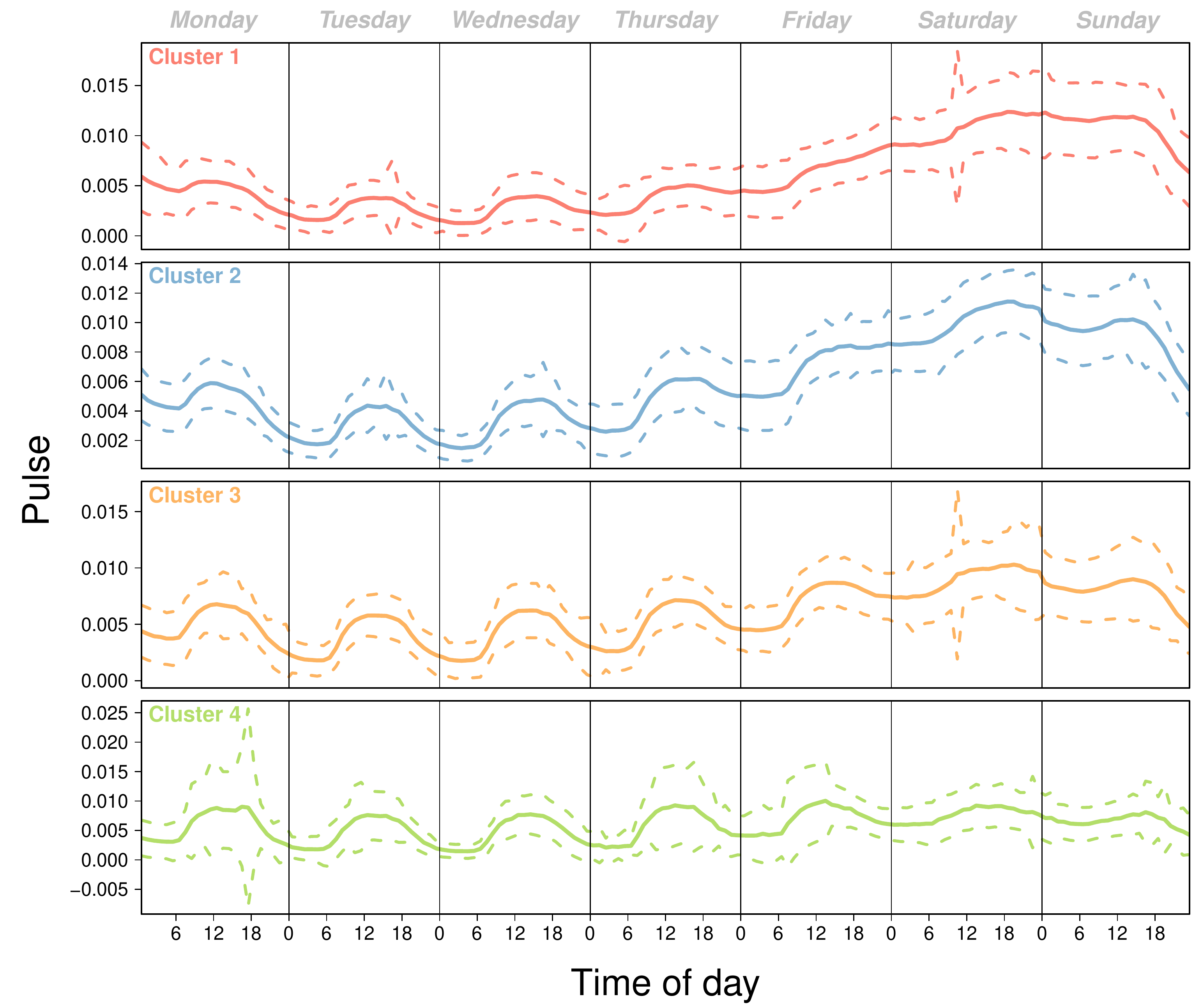}
	\caption{\textbf{Pulse associated with the four main clusters.} The solid lines represent the average pulse, while the dashed lines represent one standard deviation. \label{FigA7}}
\end{figure}

\begin{figure}[!h]
	\centering 
	\includegraphics[width=\linewidth]{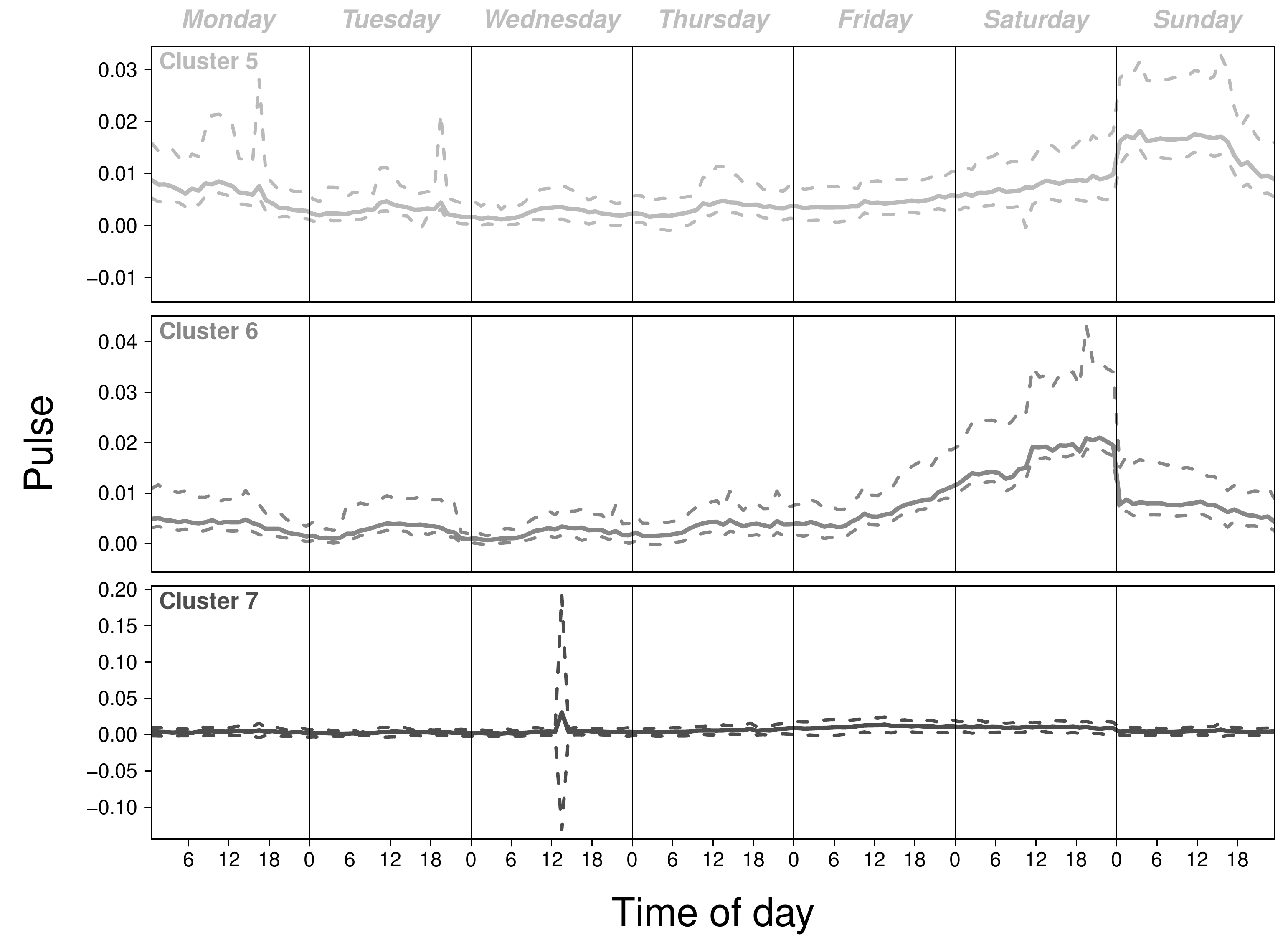}
	\caption{\textbf{Pulse associated with the three additional clusters.} The solid lines represent the average pulse, while the dashed lines represent one standard deviation. 
		\label{FigA8}}
\end{figure}

\begin{figure}[!h]
	\centering 
	\includegraphics[width=12cm]{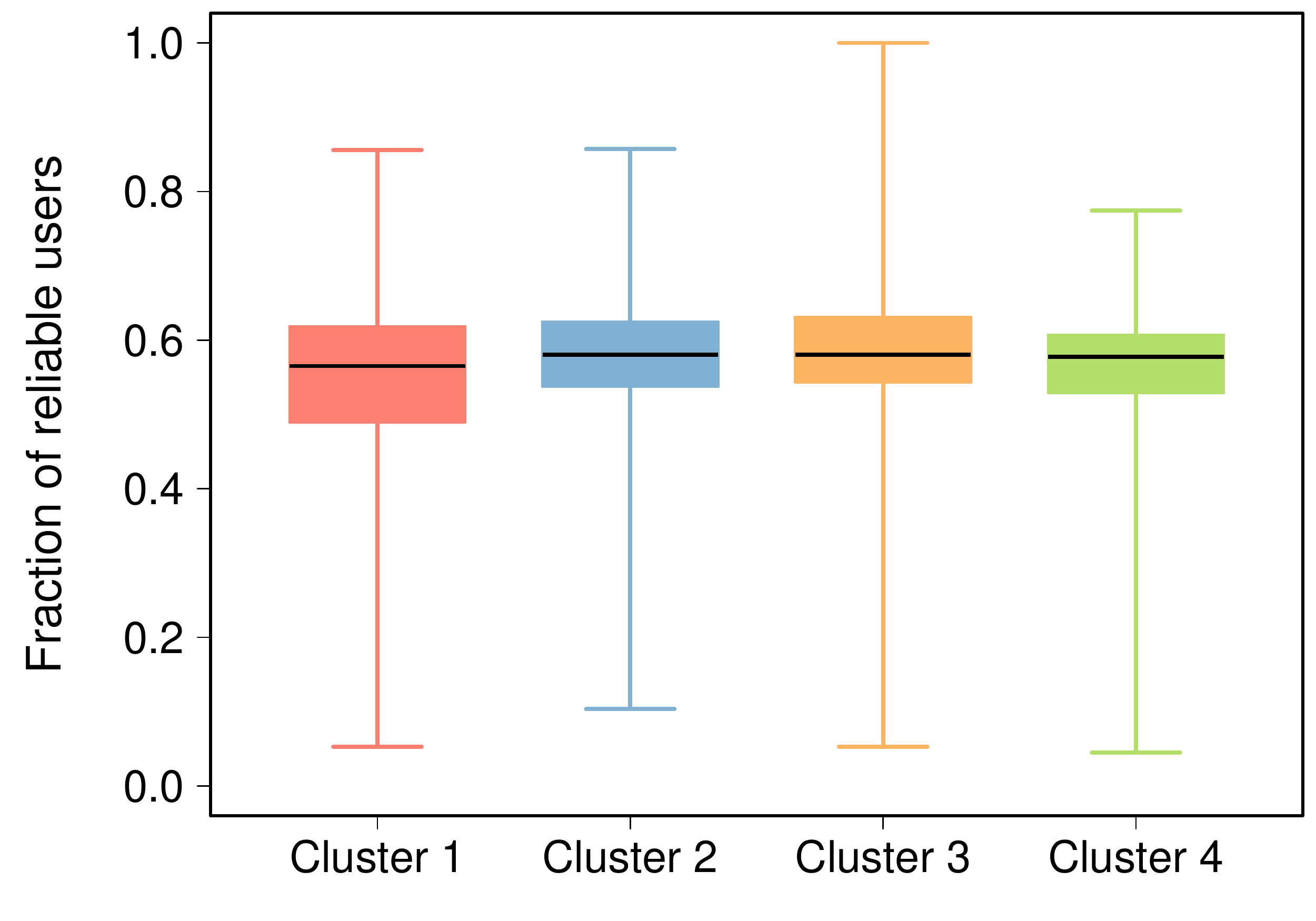}
	\caption{\textbf{Boxplots of the fraction of reliable users per cluster.}Each boxplot is composed of the minimum value, the first quartile, the median, the third quartile and the maximal value. \label{FigA9}}
\end{figure}

\clearpage
\subsection*{Null model}

\begin{figure}[!h]
	\centering 
	\includegraphics[width=12cm]{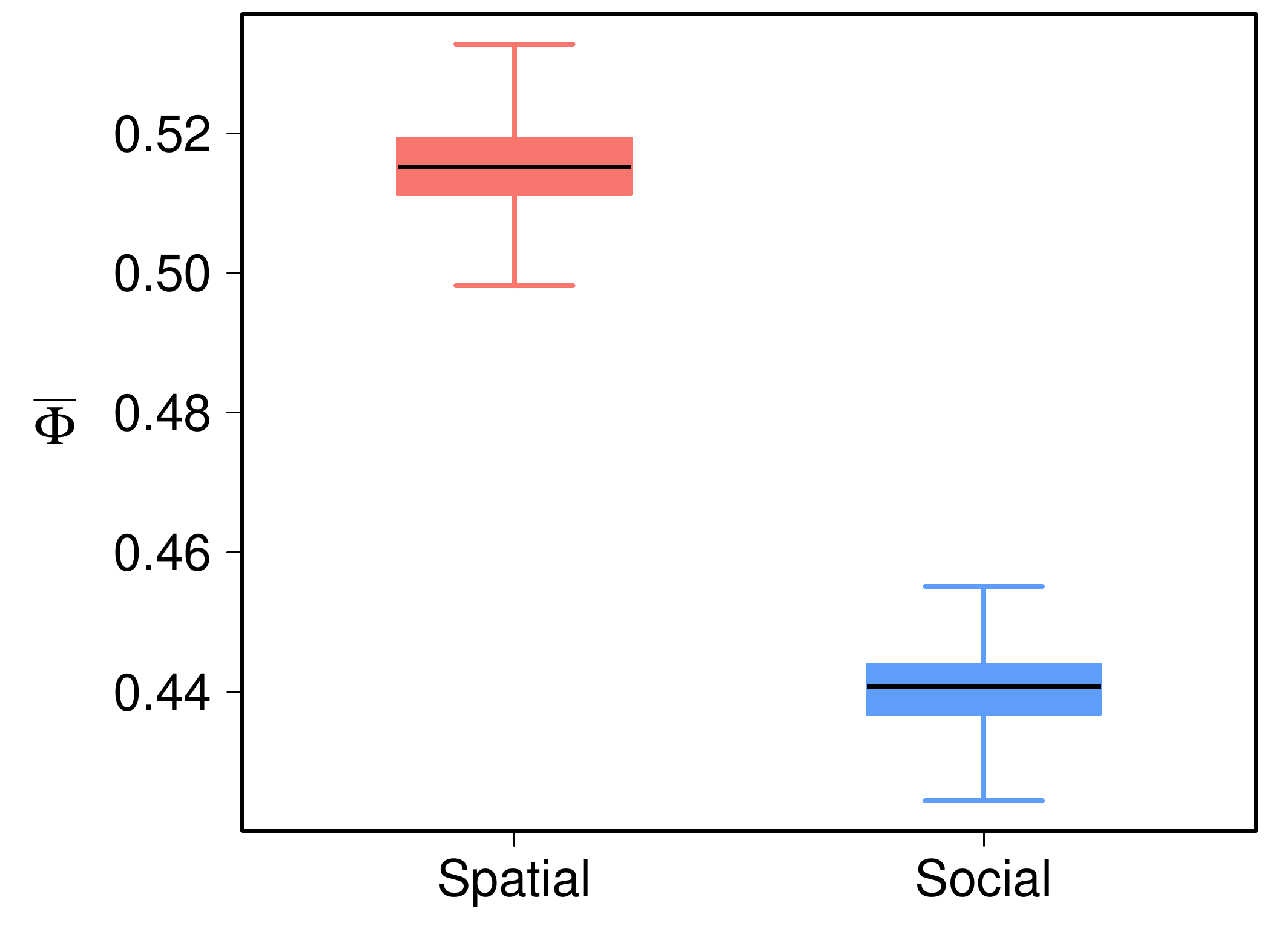}
	\caption{\textbf{Boxplots of $\bar{\Phi}$ for the spatial and social interaction matrices.} Each boxplot is composed of 100 $\bar{\Phi}$ values, each of them obtained with a $\Phi_h$ value based on one random assignment. Each boxplot is composed of the minimum value, the first quartile, the median, the third quartile and the maximal value. \label{FigA10}}
\end{figure}

\end{document}